\documentclass[a4paper,aps,pra,twocolumn,showpacs,preprintnumbers,amsmath,amssymb]{revtex4}
\usepackage{bbm}
\usepackage{amsmath}
\usepackage{verbatim}
\usepackage{graphicx}

\newcommand{\p}[3]{p_{\mathrm{\,#1},#2}^\mathrm{\,#3}}
\newcommand{\x}[3]{x_{\mathrm{\,#1},#2}^\mathrm{\,#3}}
\newcommand{\xa}[1]{X^{\,\mathrm{#1}}}
\newcommand{\pa}[1]{P^{\,\mathrm{#1}}}
\newcommand{\y}[2]{y_{\mathrm{\,#1},#2}}
\newcommand{\q}[2]{q_{\mathrm{\,#1},#2}}
\newcommand{\Eqref}[1]{Eq.~(\ref{#1})}
\newcommand{\Eqsref}[1]{Eqs.~(\ref{#1})}
\newcommand{\co}[1]{\alpha_{#1}}
\newcommand{\dB}{\,\mathrm{dB}}
\begin{document}

\title{High fidelity teleportation between light and atoms}

\author{K. Hammerer$^{1,2}$, E.S. Polzik$^{3}$, J.I. Cirac$^4$}

\affiliation{ $^1$Institute for Theoretical Physics, University of Innsbruck, Technikerstrasse, A-6020 Innsbruck, Austria \\
$^2$Institute for Quantum Optics and Quantum Information of the
Austrian Academy of Sciences, Technikerstrasse, A-6020 Innsbruck, Austria\\
$^3$Niels Bohr Institute, QUANTOP, Danish Research Foundation Center for Quantum Optics, Blegdamsvej, DK-2100 Copenhagen, Denmark\\
$^4$Max-Planck--Institut f\"ur Quantenoptik,
Hans-Kopfermann-Strasse, D-85748 Garching, Germany}

\begin{abstract}
We show how high fidelity quantum teleportation of light to atoms
can be achieved in the same setup as was used in the recent
experiment [J. Sherson et.al., quant-ph/0605095, accepted by
Nature], where such an inter-species quantum state transfer was
demonstrated for the first time. Our improved protocol takes
advantage of the rich multimode entangled structure of the state of
atoms and scattered light and requires simple post-processing of
homodyne detection signals and squeezed light in order to achieve
fidelities up to 90\% (85\%) for teleportation of coherent (qubit)
states under realistic experimental conditions. The remaining
limitation is due to atomic decoherence and light losses.
\end{abstract}

%03.67.-a Quantum information
%03.65.Ud Entanglement and quantum nonlocality (e.g. EPR paradox, Bell's inequalities, GHZ states, etc.)
%42.50.-p Quantum optics
%42.50.Dv Nonclassical field states; squeezed, antibunched, and sub-Poissonian states; operational definitions of the phase of the field; phase measurements
%32.80.-t Photon interactions with atoms (see also 42.50 Quantum optics)
%32.80.Qk Coherent control of atomic interactions with photons

\maketitle

The first quantum teleportation of light on atoms was recently
demonstrated by J. Sherson et.al. \cite{JKOJHCP}. Based on a
protocol proposed in \cite{HPC}, the experiment utilized
entanglement between a cloud of atoms and a propagating pulse to
transfer the coherent state carried by an other, independent pulse
to the collective spin state of atoms. Measured fidelities ranging
from 56\% up to 64\% clearly constitute a better-than-classical
transfer of coherent states \cite{HWPC} and essentially prove that
there indeed was entanglement of light and atoms present. If quantum
teleportation of light on atoms was to be used as a building block
of a quantum network, requirements on its performance will of course
be more stringent. For example, recent security analyses of
continuous variable quantum cryptography \cite{GC, NGA} prove that
the tolerable amount of excess noise will be in any case below 0.4
(0.8) shot noise units for protocols based on coherent (squeezed)
states, corresponding to fidelities of 83\% (71\%). In this paper we
elaborate on methods to improve the protocol of \cite{HPC, JKOJHCP}
such as to attain high fidelity teleportation of light on atoms.

This goal can be achieved in two ways: First, a simple
post-processing of homodyne detection signals recorded in the Bell
measurement followed by a suitable feedback onto atoms already
yields a significant enhancement. The strategy is based on the idea
to include in the description also higher order temporal scattering
modes, which were treated as noise in the original protocol
\cite{HPC}. In this way it is possible to benfit from the rich,
multimode entangled structure inherent to the state of scattered
light and atoms. Second, the remaining excess noise in atoms will be
due to vacuum fluctuations of light, which can be reduced by using
squeezed light already at the step of entangling light and atoms.
Combined, these two methods will yield a fidelity which is limited
by light losses and atomic decoherence only and can get close to
90\% under realistic experimental conditions for teleportation of
coherent states. We also study the teleportation of qubit states,
encoded in superpositions of vacuum and a single photon state, for
which we predict a fidelity close to 85\% under the same conditions.
In the following we will first extend the model developed in
\cite{HPC} and give a complete description of the entangled state of
light and atoms, which will be used as a resource in the
teleportation protocols presented thereafter.

\paragraph*{Resource state} The system is the same as in \cite{HPC, JKOJHCP}. An ensemble of
$N_{at}$ Alkali atoms, whose ground state spins are maximally
polarized along the $x$-direction, is immersed in a homogeneous
magnetic field aligned along the same direction. The transverse
collective spin components, Larmor precessing at a frequency
$\Omega$, can be described by canonical operators $[X,P]=i$ with
zero mean and a normalized variance \mbox{$\Delta X^2=\Delta
P^2=1/2$} for the initial coherent spin state. A strong coherent
pulse of frequency $\omega_0$ and linearly polarized along $x$ is
then sent through the atomic sample along the $z$-direction. The
scattered, $y$-polarized light is described in terms of spatially
localized modes $[x(z),p(z')]=ic\delta(z-z')$. These modes are
initially in vacuum such that \mbox{$\langle x(z)\rangle=\langle
p(z)\rangle=0$} and \mbox{$\langle x(z)x(z')\rangle=\langle
p(z)p(z')\rangle=c\delta(z-z')/2$}. The dynamics of this system can
be described by the effective Hamiltonian \cite{HPC,J,KMSJP}
\begin{align}\label{Hamiltonian}
H&=H_{at}+H_{li}+V.
\end{align}
$H_{at}=\hbar\Omega(X^2+P^2)/2$ describes the effect of the magnetic
field, $H_{li}$ the propagation of light and $V=\hbar\kappa
Pp(0)/\sqrt{T}$ the effective interaction of light and atoms. $T$ is
the pulse length and $\kappa$ a dimensionless coupling constant
given by
\mbox{$\kappa=\sqrt{N_{ph}N_{at}F}a_1\sigma\Gamma/2A\Delta$} where
$N_{ph}$ is the overall number of photons in the pulse, $a_1$ is a
constant characterizing the ground state's vector polarizability,
$\sigma$ is the scattering cross section, $\Gamma$ the decay rate,
$A$ the effective beam cross section and $\Delta$ the detuning from
the probed transition. Note that the effective form of the
interaction $V$ is true only in the case where $\Delta$ is larger
than the typical exited states' hyperfine splitting \cite{J,KMSJP}.
In the following we will study the unitary evolution according to
Hamiltonian \eqref{Hamiltonian}. Atomic dephasing and light losses
in this system are small and will be treated perturbatively as
linear losses.

Hamiltonian \eqref{Hamiltonian} gives rise to a set of linear
Maxwell-Bloch equations, whose explicit form can be found in
\cite{HPC}. These equations have to be integrated up to a time $T$,
when the strong pulse triggering the interaction leaves the sample.
This can be done along the lines of \cite{HPC} and the solution can
be phrased in terms of a set of temporal scattering modes defined as
\begin{align}\label{modedef}
\x{c}{n}{in}=\frac{\mathcal{N}_n}{\sqrt{T}}\int_0^T\!\!d\tau\cos(\Omega\tau)\bar{P}_n(\tau/T)\bar{x}(c\tau,0).
\end{align}
Here $n=0,1,2,\ldots$, $\bar{P}_n(x)=P_n(2x-1)$, where $P_n(x)$ is
the n-th Legendre polynomial and $\mathcal{N}_n$ is a normalization
constant. Analogous definitions hold for $\p{c}{n}{in}$ with
$\bar{x}$ replaced by $\bar{p}$ and for
$\x{s}{n}{in},\,\p{s}{n}{in}$ with $\cos(\Omega\tau)$ replaced by
$\sin(\Omega\tau)$. In the limit of $\Omega T\gg 1$, which is well
fulfilled under usual experimental conditions \cite{JKOJHCP} where
$T$ is on the order of several $\mathrm{ms}$ and $\Omega$ of some
$100\,\mathrm{kHz}$, and for $n\ll\Omega T$ these modes are
effectively orthogonal,
\mbox{$[\x{\alpha}{n}{},\p{\beta}{m}{}]=i\delta_{\alpha\beta}\delta_{n,m}\,(\alpha,\beta=\mathrm{c,s})$},
and the normalization is given by $\mathcal{N}_n=\sqrt{4n+2}$. By
means of these modes the final state of atoms and scattered light
can be expressed as
\begin{subequations}\label{outstate}
\begin{align}
\xa{out}&=\xa{in}+\frac{\kappa}{\sqrt{2}}\p{c}{0}{in}\label{Xout}\\
\pa{out}&=\pa{in}+\frac{\kappa}{\sqrt{2}}\p{s}{0}{in}\label{Pout}\\
\p{\alpha}{n}{out}&=\p{\alpha}{n}{in}\quad
(n\geq0,\,\alpha=\mathrm{c,s}),\\
\x{c}{0}{out}&=\x{c}{0}{in}+\frac{\kappa}{\sqrt{2}}\pa{in}+\left(\frac{\kappa}{2}\right)^2\left(\p{s}{0}{in}-\frac{1}{\sqrt{3}}\p{s}{1}{in}\right),\label{xc0}\\
\x{s}{0}{out}&=\x{s}{0}{in}-\frac{\kappa}{\sqrt{2}}\xa{in}-\left(\frac{\kappa}{2}\right)^2\left(\p{c}{0}{in}-\frac{1}{\sqrt{3}}\p{c}{1}{in}\right),\label{xs0}\\
\x{c}{n}{out}&=\x{c}{n}{in}+\left(\frac{\kappa}{2}\right)^2\left(\alpha_n\p{s}{n-1}{in}-\alpha_{n+1}\p{s}{n+1}{in}\right),\\
\x{s}{n}{out}&=\x{s}{n}{in}-\left(\frac{\kappa}{2}\right)^2\left(\alpha_n\p{c}{n-1}{in}-\alpha_{n+1}\p{c}{n+1}{in}\right),\label{xsn}
\end{align}
\end{subequations}
where $\alpha_n=1/\sqrt{4n^2-1}$ and the last two equations concern
the cases $n\geq1$ only. Obviously correlations between atoms and
the zero order light modes are created proportional to $\kappa$. In
addition, proportional to $\kappa^2$, there are also correlations
induced between the various higher order light modes. This
back-action effect of light onto itself was clearly visible in the
measurements of $\x{\alpha}{0}{}$ performed in \cite{JKOJHCP} and
were well described by \Eqref{xc0} and \eqref{xs0}.

As opposed to \cite{HPC,JKOJHCP}, where only the atomic mode and the
$n=0$ light modes were considered as being part of the system, our
aim here is to take advantage also of the correlations created among
the higher order scattering modes in order to improve the
teleportation fidelity. The protocol proceeds as follows:
\paragraph*{Input state} The quantum state to be teleported is
encoded in a mode $[y,q]=i$ given by
\begin{align}\label{InputMode}
y&=\sum_{n=0}^{N}\frac{c_n}{\sqrt{2}}\left(\y{s}{n}+\q{c}{n}\right),
&q&=\sum_{n=0}^{N}\frac{c_n}{\sqrt{2}}\left(\q{s}{n}-\y{c}{n}\right),
\end{align}
where the modes
$[\y{\alpha}{n},\q{\beta}{m}]=i\delta_{\alpha,\beta}\delta_{n,m}$
are defined analogously to \Eqref{modedef} and the coefficients
$c_n$ are real and fulfill $\sum_n c_n^2=1$. This mode is centered
in frequency at the upper sideband $\omega_0+\Omega$ and has a
slowly varying envelope determined by the coefficients $c_n$ and
$N$, which both will be specified later. The quantum state of this
mode can in principle be arbitrary, but we will focus on coherent
states in the following.
\paragraph*{Bell measurement} The scattered light in
$y$-polarization, described by \Eqsref{xc0} to \eqref{xsn},
interferes at a balanced beam splitter with the input field. After
the beam splitter the commuting observables
\begin{align}
\tilde{x}_{\mathrm{c},n}&=\frac{1}{\sqrt{2}}\left(\x{c}{n}{out}+\y{c}{n}\right),
&\tilde{x}_{\mathrm{s},n}&=\frac{1}{\sqrt{2}}\left(\x{s}{n}{out}+\y{s}{n}\right),\nonumber\\
\tilde{q}_{\mathrm{c},n}&=\frac{1}{\sqrt{2}}\left(\p{c}{n}{out}-\q{c}{n}\right),
&\tilde{q}_{\mathrm{s},n}&=\frac{1}{\sqrt{2}}\left(\p{s}{n}{out}-\q{s}{n}\right)\label{MeasuredModes}
\end{align}
are measured up to $n_{max}$. This can be achieved by multiplying
the photocurrent resulting from a standard polarimetric measurement
of Stokes vector components $S_y$ (or $S_z$) with the pulse
envelopes given in \Eqref{modedef} and integrating over the pulse
duration.
\paragraph*{Feedback} Let the measurement outcomes corresponding to
the observables above be given by
$\tilde{X}_{\mathrm{c},n},\,\tilde{X}_{\mathrm{s},n},\,\tilde{Q}_{\mathrm{c},n}$
and $\tilde{Q}_{\mathrm{s},n}$ respectively. The atomic state is
then displaced by an amount
\mbox{$\sum_nc_n(\tilde{X}_{\mathrm{s},n}-\tilde{Q}_{\mathrm{c},n})$}
in $X$ and
\mbox{$-\sum_nc_n(\tilde{X}_{\mathrm{c},n}+\tilde{Q}_{\mathrm{s},n})$}
in $P$. In the ensemble average the final atomic state is then given
by
$\xa{fin}=\xa{out}+\sum_nc_n(\tilde{x}_{\mathrm{s},n}-\tilde{q}_{\mathrm{c},n})$
and
$\pa{fin}=\pa{out}-\sum_nc_n(\tilde{x}_{\mathrm{c},n}+\tilde{q}_{\mathrm{s},n})$,
such that, by means of \Eqsref{outstate},\eqref{InputMode} and
\eqref{MeasuredModes} we arrive at
\begin{align}\label{finalstate}
\xa{fin}&=y+\frac{1}{\sqrt{2}}\sum_{n=0}^N c_n\x{s}{n}{in}
+\left(1\!-\!\frac{c_0\kappa}{2}\right)\xa{in}-\!\!\sum_{n=0}^{N+1}f_n\p{c}{n}{in},\nonumber\\
\pa{fin} &=q-\frac{1}{\sqrt{2}}\sum_{n=0}^N c_n\x{c}{n}{in}
+\left(1\!-\!\frac{c_0\kappa}{2}\right)\pa{in}-\!\!\sum_{n=0}^{N+1}f_n\p{s}{n}{in},
\end{align}
where the coefficients $f_n$ are
\begin{align*}
f_0&=\frac{1}{\sqrt{2}}\left[c_0-\kappa+\left(\frac{\kappa}{2}\right)^2\left(c_0+c_1\co{1}\right)\right],\\
f_n&=\frac{1}{\sqrt{2}}\left[c_n+\left(\frac{\kappa}{2}\right)^2\left(c_{n+1}\co{n+1}-c_{n-1}\co{n}\right)\right]\quad(n\geq1)
\end{align*}
and one has to set $c_{N+1}=c_{N+2}=0$ in the sums in
\eqref{finalstate}. As is evident from \Eqsref{finalstate}, atoms
receive the correct light mode (first terms on the r.h.s.) as well
as a certain amount of excess noise (remaining three terms).

\begin{figure}[t]\begin{center}
\includegraphics[width=8.5cm]{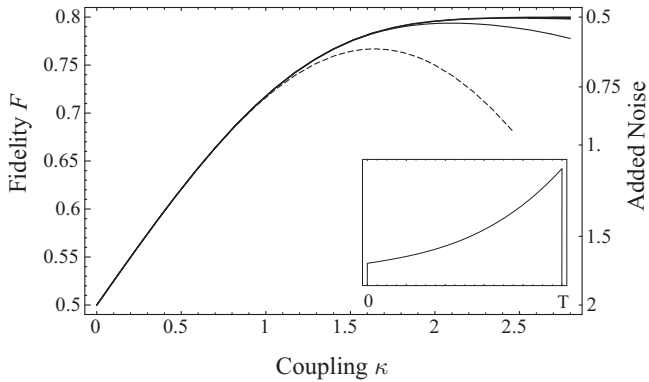}
\caption{Teleportation fidelity $F$ and added noise in units of shot
noise for coherent states versus coupling $\kappa$ for different
numbers of modes $N$ included in the protocol, optimized for the
parameters $c_n$ determining the pulse shape of the input mode. The
dashed line is the fidelity achievable if only the zeroth order
mode, $N=0$, is included (c.f. \cite{HPC,JKOJHCP}). The thin line
corresponds to $N=1$. The results for $N=2$ and $N=3$ (thick line)
are practically identical and saturate the bound of $F=80\%$. Inset:
Optimal envelope for the input field for the case $N=3,\,\kappa=2$,
normalized over pulse duration $[0,T]$.}\label{Fidelities0}
\end{center}\end{figure}

\begin{figure}[t]\begin{center}
\includegraphics[width=8.5cm]{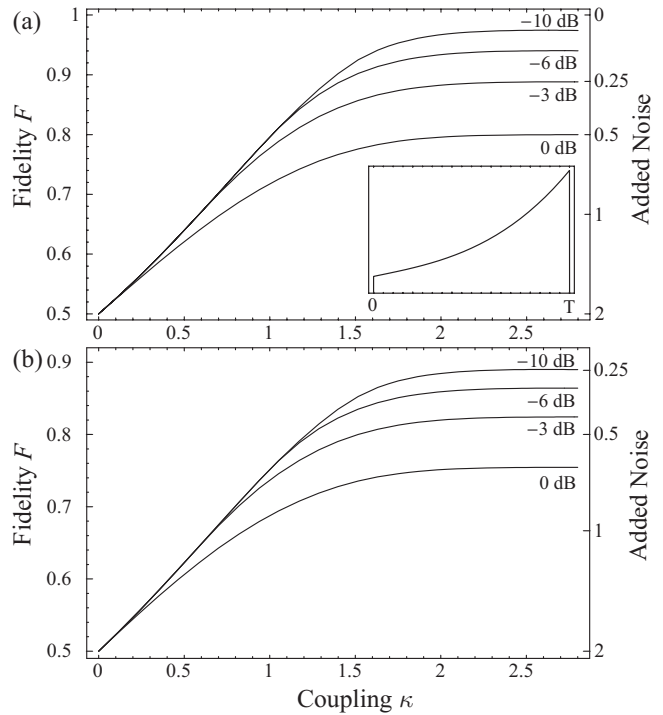}
\caption{(a) Teleportation fidelity $F$ and added noise in units of
shot noise for coherent states versus coupling $\kappa$ for
protocols involving four modes ($N=3$) using squeezing of light down
to $-3\dB,-6\dB$ and $-10\dB$. The curve for $0\dB$ is identical to
the thick line of Figure \ref{Fidelities0}. Inset: Optimal envelope
for the input field for $-10\dB$ of
squeezing and $\kappa=2$, normalized over pulse duration $[0,T]$.\\
(b) Teleportation fidelity $F$ and added noise in units of shot
noise for coherent states versus coupling $\kappa$ including 10\%
atomic decay and 10\% light losses. The optimal pulse shape is
practically identical to the one shown in the inset of
(a).}\label{Fidelities1}
\end{center}\end{figure}

For unit-gain teleportation of coherent states, it is the variance
of the latter terms which limits the teleportation fidelity (see
\cite{HPC} for a definition) and therefore has to be minimized - for
a given coupling $\kappa$ - by a proper choice of the coefficients
$c_n$. Respecting the normalization condition $\sum_nc_n^2=1$ the
best result that can be expected from such a strategy would be a
cancelation of the last two terms in both of \Eqsref{finalstate}. In
this case the final atomic state would be $\xa{fin}=y+\sum_n
c_n\x{s}{n}{in}/\sqrt{2}$ and $\pa{fin}=q-\sum_n
c_n\x{c}{n}{in}/\sqrt{2}$, which amounts to half a unit of vacuum
noise added to both spin components or a fidelity of $80\%$. Figure
\ref{Fidelities0} shows the result of such an optimization for
different choices of $N$, that is the number of modes which are
included in the protocol. The limiting value of $80\%$ can in deed
be achieved by taking into account the first three higher order
modes only.
In order to beat also this limit, observe first that the half unit
of vacuum noise added to the atomic state is due to the initial
vacuum noise of modes $\x{\alpha}{n}{in}$, that is the vacuum field
in $y$-polarization copropagating with the classical $x$-polarized
pump field. These vacuum fluctuations can be suppressed by injecting
squeezed light along with the classical field. The squeezing
spectrum has to be broad enough such as to cover the sidebands at
$\pm\Omega$ which is readily provided by a state of the art source
of squeezed light, whose squeezing spectrum typically covers several
MHz. Using squeezed light, the final atomic variance is
$(\Delta\xa{fin})^2=(\Delta\pa{fin})^2=1+s/2$, where $s$ is the
squeezed variance of $y$-polarized light and the corresponding
fidelity is $F=2/(2+s/2)$, ranging from $80\%$ for $s=0$ (no
squeezing) approaching $100\%$ for $s\rightarrow 0$. Figure
\ref{Fidelities1} shows the result of protocols involving four
temporal modes ($N=3$) and vacuum noise reduction down to
$s=0.5,\,s=0.25$ and $s=0.1$ corresponding to about
\mbox{$-3\dB,-6\dB$} and $-10\dB$ of light squeezing respectively.
Fidelities level off at the values expected from the simple formula
given above and are thus bounded by the amount of single-mode
squeezing, which reminds of the situation for continuous variable
light-to-light teleportation \cite{V, BK}, whose performance is
limited by the amount of two-mode squeezing.
\paragraph*{Losses} Up to this point we have neglected any effects of decoherence,
which will inevitably occur due to spontaneous emission and
absorption of light. As discussed in \cite{HPC}, these processes can
be treated as linear losses, such that f.e. the state of atoms after
the scattering is given by
\begin{align*}
\xa{out}&=\sqrt{1-\beta}\left(\xa{in}+\frac{\kappa}{\sqrt{2}}\p{c}{0}{in}\right)+\sqrt{\beta} f_X,\\
\pa{out}&=\sqrt{1-\beta}\left(\pa{in}+\frac{\kappa}{\sqrt{2}}\p{s}{0}{in}\right)+\sqrt{\beta} f_P,\\
\end{align*}
rather than by \Eqsref{Xout} and \eqref{Pout}. Passive light losses
are naturally described by similar expressions for
$\x{\alpha}{n}{out},\p{\alpha}{n}{out}$. We assume that all light
modes of interest are affected by the same amount of $(1-\epsilon)$
of power loss. Simple considerations show that in this case we can
expect a fidelity of $F=2/(2+s/2+\beta+\epsilon)$. Figure
\ref{Fidelities1} shows the numerical results including $\beta=10\%$
decay of transverse spin components and $\epsilon=10\%$ absorption
losses in each light mode, which corresponds to the experimental
conditions of \cite{JKOJHCP}. The simple bound given above agrees
again well with numerical results.
\paragraph*{Non-unit gain teleportation} So far we considered only unit-gain teleportation
of coherent states, that is, we required that amplitudes are
transmitted faithfully. If however it is known that the coherent
states to be teleported are drawn from a certain pre-defined set
only, such as a Gaussian distribution around the vacuum state, it
might be advantageous to accept a certain mismatch in amplitude in
order to reduce the added noise. The protocol described above is
easily generalized to non-unit gain feedback by simply replacing the
feedback coefficients $c_n$ in \Eqsref{finalstate} by $gc_n$, where
$g$ is now a suitably chosen gain. The result will be
\begin{align}
\langle\xa{fin}\rangle&=g\langle y\rangle&(\Delta\xa{fin})^2&=g^2\Delta y^2+\Delta F^2,\nonumber\\
\langle\pa{fin}\rangle&=g\langle
q\rangle&(\Delta\pa{fin})^2&=g^2\Delta q^2+\Delta
F^2,\label{MeansVars}
\end{align}
where $\Delta F^2$ represents the variance of the last three terms
in \Eqsref{finalstate}. From this one can evaluate the teleportation
fidelity, averaged over the set of input states, and optimize for
$g$. As an example, for a Gaussian distributed set of input states
of mean photon number $\bar{n}=2$, the average fidelity can be 90\%
for $\kappa=1.5$ and light squeezing of 10 dB including 10\% atomic
dephasing and 10\% light losses. The optimal gain in this case is
$g=0.9$.
\begin{figure}[t]\begin{center}
\includegraphics[width=7.5cm]{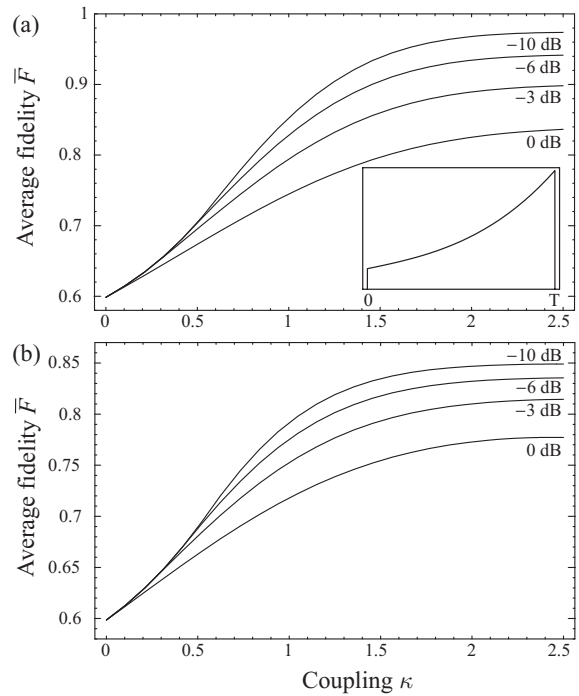}
\caption{(a) Teleportation fidelity $F$ for qubit states versus
coupling $\kappa$ for protocols involving four modes ($N=3$) using
squeezing of light down to $-3\dB,-6\dB$ and $-10\dB$, averaged over
the Bloch sphere. Inset: Optimal envelope for the input field for
$-10\dB$ of
squeezing and $\kappa=2$, normalized over pulse duration $[0,T]$.\\
(b) Teleportation fidelity $F$ for qubit states versus coupling
$\kappa$ including 10\% atomic decay and 10\% light losses, averaged
over the Bloch sphere. The optimal pulse shape is practically
identical to the one in shown in the inset of
(a).}\label{Fidelities4}
\end{center}\end{figure}
\paragraph*{Teleportation of qubit states} Having discussed the teleportation
of coherent states, it is interesting to ask how well the proposed
protocol would work, if the state to be teleported was a
non-gaussian state, f.e. a qubit-state
$|\psi(\theta,\phi)\rangle=\cos(\theta/2)|0\rangle+\exp(i\phi)\sin(\theta/2)|1\rangle$,
where $|0\rangle$ and $|1\rangle$ are the vacuum and the
single-photon state of mode \eqref{InputMode} respectively. It is
clear that unit fidelity of coherent state teleportation implies
unit fidelity for the teleportation of arbitrary states, as coherent
states provide a basis in Hilbert space. But also for imperfect
protocols one can expect a rather high teleportation fidelity for
states involving only few photons given the result of the previous
paragraph. In order to explicitly evaluate the fidelity for states
of the form $|\psi(\theta,\phi)\rangle$, note that the whole
teleportation protocol implements a completely positive map
$\mathcal{E}$ such that the (density operator of the) state of atoms
after the teleportation protocol is
$\mathcal{E}(|\psi\rangle\langle\psi|)$ when $|\psi\rangle$ was the
input. The map $\mathcal{E}$ is a general Bogoliubov transformation
and fixed by the linear input-output relations \eqref{finalstate}.
Evaluating the corresponding fidelity,
$F=\langle\psi|\mathcal{E}(|\psi\rangle\langle\psi|)|\psi\rangle$,
is not straight forward, as $\mathcal{E}$ - loosely speaking - mixes
creation and annihilation operators. In order to calculate the
fidelity $F$ one can take advantage of (\textit{i}) relation
\begin{equation*}
|n\rangle\langle
m|=\left[\frac{\partial^{n+m}}{\partial\alpha^n\partial{\alpha^*}^m}\left(e^{\alpha\alpha^*}|\alpha\rangle\langle\alpha|\right)\right]\!\!\Bigg|_{\alpha=\alpha^*=0}\!\!\!\!,\,
(n,m=0,1),
\end{equation*}
where $|n\rangle$ is a Fock state and $|\alpha\rangle$ a coherent
state and (\textit{ii}) the fact that the action of $\mathcal{E}$ on
coherent states has the simple representation
\begin{equation*}
\mathcal{E}(|\alpha\rangle\langle\alpha|)=\frac{1}{2\pi\sigma^2}\int
d^2\beta e^{-|\beta-g\alpha|^2/2\sigma^2}|\beta\rangle\langle\beta|,
\end{equation*}
where $\sigma^2$ is given by the variance of atomic spin components
c.f. \Eqsref{MeansVars}, that is
$\sigma^2=[(\Delta\xa{fin})^2-1]/4=[(\Delta\pa{fin})^2-1]/4$ with
variances measured in units of vacuum noise. By means of these
relations the fidelity, averaged over the Bloch sphere, is found to
be
\begin{align}\label{qubitfidel}
\bar{F}&=\frac{1}{4\pi}\int
d\Omega\langle\psi|\mathcal{E}(|\psi\rangle\langle\psi|)|\psi\rangle\nonumber\\
&=\frac{3+2g+g^2+2(9+2g-3g^2)\sigma^2+24\sigma^4}{6(1+2\sigma^2)^3}.
\end{align}
This expression can now again be optimized with respect to the gain
$g$ and the input-envelope fixed by $c_n$. The results are shown in
Figure \ref{Fidelities4} and prove that it is well possible to
violate the classical benchmark of $2/3$ for the teleportation of
qubit states. Note that relation \eqref{qubitfidel} holds for all
Gaussian maps of the form \eqref{MeansVars} and is thus of relevance
also in other situations such as for example for evaluating the
efficiency of quantum memory protocols \cite{MM, FSOP, SSFMP, MHPC}.

In summary we showed that the protocol used in \cite{JKOJHCP} to
perform teleportation of light on atoms can be improved to yield
high fidelities up to 90\% under realistic conditions. The final
limitation comes from decoherence of atoms and light losses which
both are on the 10\% level in the setup of \cite{JKOJHCP}, which is
a room-temperature ensemble of atoms in a glass cell. In particular
in the balance of light losses 5\% are due to propagation losses and
detector inefficiencies and 5\% come from reflections from walls of
the glass cell. Losses of the latter kind can be reduced down to
0.5\% with improved anti-reflection coating. Furthermore, we expect
that for cold trapped atoms, eventually in an optical lattice, both
atomic dephasing and light losses can be diminished significantly.
For a different proposal of light-to-atom teleportation based on
collective recoil in a Bose-Einstein condensate see \cite{PCPB}

We acknowledge funding from EU projects COVAQIAL, QAP, CONQUEST and
SCALA and from the Austrian Science Foundation.


\begin{thebibliography}{54}


\bibitem{JKOJHCP} J. Sherson, H. Krauter, R.K. Olsson, B. Julsgaard, K. Hammerer, I. Cirac, E.S.
Polzik, quant-ph/0605095

\bibitem{HPC} K. Hammerer, E.S. Polzik, J.I. Cirac, Phys. Rev. A \textbf{72}, 052313 (2005)

\bibitem{HWPC} K. Hammerer, M.M. Wolf, E.S. Polzik, J.I. Cirac, Phys. Rev. Lett. \textbf{94}, 150503 (2005)

\bibitem{GC} R. Garcia-Patron, N.J. Cerf, quant-ph/0608032

\bibitem{NGA} M. Navascues, F. Grosshans, A. Acin, quant-ph/0608034

\bibitem{J} B. Julsgaard, \textit{Entanglement and Quantum
Interactions with Macroscopic Gas Samples}, PhD Thesis, October
2003, http://www.nbi.dk/$\sim$julsgard/

\bibitem{KMSJP} D.V. Kupriyanov, O.S. Mishina, I.M. Sokolov, B. Julsgaard, and E.S. Polzik
Phys. Rev. A 71, 032348 (2005), O. Mishina, D. Kupriyanov, E.S.
Polzik, quant-ph/0509220


\bibitem{V} L. Vaidman, Phys. Rev. A, \textbf{49}, 1473 (1994)

\bibitem{BK} S.L. Braunstein, H.J. Kimble, Phys. Rev. Lett. \textbf{80}, 869
(1998)

\bibitem{MM}  L.B. Madsen, K. Molmer, Phys. Rev. A \textbf{73}, 032342
(2006)

\bibitem{FSOP} J. Fiurasek, J. Sherson, T. Opatrny, E.S. Polzik, Phys.
Rev. A \textbf{73}, 022331 (2006)

\bibitem{SSFMP} J. Sherson, A.S. Soerensen, J. Fiurasek, K. Moelmer, E.S. Polzik, quant-ph/0505170

\bibitem{MHPC} C.A. Muschik, K. Hammerer, E.S. Polzik, J.I. Cirac, Phys. Rev. A \textbf{73}, 062329
(2006)

\bibitem{PCPB} M.G.A. Paris, M. Cola, N. Piovella, R. Bonifacio, Opt. Commun. {\bf 227}, 349
(2003)

\end{thebibliography}
\end{document}